# Dynamics of ranking processes in small-world networks

**Bahruz Gadjiev, Tatiana Progulova, Sergey Timoshin**

*International University for Nature, Society and Man*

**Abstract.** In the paper we discuss the dynamics of the order parameter in complex networks with long-range space interactions and temporal memory and analyze phase transitions induced by noise in such systems.

## 1. Introduction

Universal critical behavior in networks is determined by the structure of a network and the symmetry of a given model. Many real networks exhibiting a small-world property, are scale-free [1, 2, 3]. The problem of studying dynamic processes in networks with such universal properties raises a question: do there exist common patterns in the dynamics of various complex systems? Further, we discuss the dynamics of the order parameter in the complex networks with small-world property and analyze phase transitions in such systems.

## 2. An equation of motion for the order parameter in the system with long-range space interactions and temporal memory

To derive an equation of motion for the order parameter in the system with long-range space interactions and temporal memory we determine the free energy functional $F[\eta]$ in the form: $F[\eta] = F_0[\eta] + F_I[\eta]$, where

$$F_0[\eta] = \int_R dr dt \int_R dr' dt' \left\{ \frac{1}{2} \frac{\partial \eta(r,t)}{\partial t} g_0(r,t,r',t') \frac{\partial \eta(r',t')}{\partial t'} \right.$$

$$\left. + \frac{1}{2} \frac{\partial \eta(r,t)}{\partial r} g_1(r,t,r',t') \frac{\partial \eta(r',t')}{\partial r'} \right\}$$

and $F_I[\eta] = \int_R dr dt \int_R dr' dt' V(\eta(r,t), \eta(r',t'))$. Here $r$ is a spatial coordinate, $t$ is time and the functions $g_0(r,t,r',t')$ and $g_1(r,t,r',t')$ describe the influence of long-range space interactions and temporal memory on the dynamics of the order parameter. Integration is performed over the region $R$ in the two-dimensional space $R^2$ to which $(r,t)$ belong. The dynamic equation follows from the stationary principle $\delta F[\eta, u] = 0$

$$\int_0^\infty dt' g(t,t') \frac{\partial \eta(t',r)}{\partial t'} + \int_0^\infty dr' k(r,r') \frac{\partial \eta(t,r')}{\partial r'}$$

$$+ \int_0^\infty d\eta(t',r') P(\eta(t,r), \eta(t',r')) \frac{\partial U(\eta(t,r), \eta(t',r'))}{\partial \eta(t',r')} = 0$$

with separated spatial and temporal kernels, where $\delta F[\eta, u]$ is the Gateaux derivative. Considering the power-like kernels $P(\eta(t,r) - \eta(t',r'))$, $g(t-t')$ and $k(r-r')$, we obtain a fractional differential equation for the order parameter as

$$g_0 {}_0^C D_t^\beta \eta(t,r) + k_0 {}_0^C D_r^\nu \eta(t,r) + {}_0^C D_{\eta(t,r)}^\mu U(\eta(t,r)) = 0 \qquad (1)$$

where ${}_0^C D_t^\beta$ is the fractional Caputo derivative. The term ${}_0^C D_{\eta(t,r)}^\mu U(\eta(t,r))$ consists of non-integer powers of the order parameter, and is a generalization of the Ginzburg – Landau equation.

3. **Noise-induced phase transitions**

In case when $k_0 = 0$ and $\beta = 1$, taking into account the influence of the external field $h\eta$ and of the multiplicative noise $\eta^s \xi(t)$ as well as of the additive noise $\varsigma(t)$ on the dynamics of the system we obtain the stochastic differential equation $\frac{\partial}{\partial t}\eta(t) = h\eta - \gamma\eta^r + \eta^s \xi(t) + \varsigma(t)$, where $\xi(t)$ and $\varsigma(t)$ are uncorrelated and the Gaussian distributed zero-mean white noise hence satisfying $\langle \xi(t)\xi(t')\rangle = 2M\delta(t-t')$ and $\langle \varsigma(t)\varsigma(t')\rangle = 2A\delta(t-t')$.

The Fokker – Plank equation for the probability function $P(\eta, t)$ associated with equation (1) can be obtained from the Kramers – Moyal expansion. Using the Stratonovich definition of the stochastic integral the Fokker – Plank equation can be written as

$$\frac{\partial P(\eta,t)}{\partial t} = -\frac{\partial}{\partial \eta}(h\eta - \gamma\eta^r)P(\eta,t) + M\frac{\partial}{\partial \eta}\left(\eta^s \frac{\partial}{\partial \eta}(\eta^s P(\eta,t))\right) + A\frac{\partial^2 P(\eta,t)}{\partial \eta^2}. \qquad (2)$$

Note that as the Caputo derivative of the constant is equal to zero the stationary solutions of this equation and of the corresponding Fokker – Planck equation with a fractional derivative coincide. Assuming that the system evolves a steady state with a time-independent $P_s(\eta)$ and a constant value for $h(t) = h_0$ we obtain the stationary solution $P_s(\eta)$ of the Fokker – Plank equation in the form $P_s(\eta) = \frac{P_0}{(A+M\eta^{2s})^{1/2}} e^{-\psi(\eta)}$ with $P_0$ being a normalization constant and $\psi(\eta) = \frac{\gamma\eta^{1+r}}{A(1+r)} {}_2F_1\left(\frac{1+r}{2s}, 1; 1+\frac{1+r}{2s}; -\frac{M}{A}\eta^{2s}\right) - \frac{h_0\eta^2}{2A} {}_2F_1\left(\frac{1}{s}, 1; 1+\frac{1}{s}; -\frac{M}{A}\eta^{2s}\right)$, where ${}_2F_1(...)$ is a hypergeometric function. To find the most probable value of $\eta$, we set the derivative of $P_s(\eta)$ equal to zero and obtain $\frac{sM\eta^{2s-1}}{A+M\eta^{2s}} + \frac{\partial\psi(\eta)}{\partial \eta} = 0$.

In the absence of multiplicative noise i.e. $M = 0$ taking into account the properties of a hypergeometric function we obtain $\frac{\partial\psi(\eta,M=0)}{\partial \eta} = 0$ and consequently $\gamma\eta^r - h_0\eta = 0$. The values of the order parameter $\eta = 0$ и $\eta = \left(\frac{\gamma}{h_0}\right)^{\frac{1}{1-r}}$ correspond to various stationary states of the system.

Consider the case when an additive noise is absent i.e. $A = 0$. Then a distribution function has the form: $P_s(\eta) = \frac{P_0}{(M\eta^{2s})^{1/2}} e^{-\varphi(\eta)}$, where $\varphi(\eta) = \frac{\gamma \eta^{1+r-2s}}{M(1+r-2s)} - \frac{h_0 \eta^{2-2s}}{M(2-2s)}$ and the stationary states of the system are determined by the equation $s + \eta \frac{\partial \varphi(\eta)}{\partial \eta} = 0$ or $sM\eta^{2s} + \gamma\eta^r - h\eta = 0$. In this case the stationary states of the system correspond to the values of the order parameter $\eta_1 = 0$ and $\eta_2 = \left(\frac{\gamma}{h_0}\right)^{\frac{1}{1-r}} + \epsilon$ where $\epsilon$ is some combination of the parameters $M, s, r, \gamma$ and $h$. Note that the phase corresponding to $\eta_2$ is induced by noise.

## 4. Non-stationary distribution function

Drift and diffusion coefficients in the Fokker-Planck equation (2) are independent of time (time-homogeneous process). As the equation is a linear one we search for a solution in the form of the product of two functions

$$P(x,t) = T(t)P_0(u) \tag{3}$$

Substitution of (3) into the equation (2) gives

$$\frac{1}{T}\frac{dT}{dt} = \frac{1}{P_0(u)} \frac{d}{du}\left[-\gamma u^s P_0(u) + Mu^s \frac{d}{du}(u^s P_0(u))\right] \tag{4}$$

The left-hand side of the equality depends only on $t$, while the right-hand one depends only on $u$. So both sides are equal to one and the same constant which we denote by $-\alpha^2$. Thus, we obtain two ordinary differential equations

$$\frac{dT(t)}{dt} = -\alpha^2 T(t) \tag{5}$$

$$\frac{d}{du}\left[-\gamma u^s P_0(u) + Mu^s \frac{d}{du}(u^s P_0(u))\right] = -\alpha^2 P_0(u) \tag{6}$$

After the change of variables $u^s P_0(u) = G(u)$ and $z = u^{-s+1}$ equation (6) is transformed to a linear second-order differential equation with constant coefficients

$$M(-s+1)\frac{d^2 G(z)}{dz^2} - \gamma \frac{dG(z)}{dz} + \frac{\alpha^2 G(z)}{-s+1} = 0 \tag{7}$$

The substitution $g(z) = G(z)e^{-\frac{1}{2M(-s+1)}\gamma z}$ leads equation (7) to the form

$$\frac{d^2 g(z)}{dz^2} - Ig(z) = 0, \tag{8}$$

where $I = -\frac{1}{M^2(-s+1)^2}\left(\frac{1}{4}\gamma^2 - \alpha^2\right)$ \hfill (9)

The solution of equation (5) has the form

$$T(t) = T_0 e^{-\alpha^2 t} \tag{10}$$

The solution of equation (6) has the form

$$g(z) = g_0 e^{k(\alpha)}, \tag{11}$$

where $k(\alpha) = \frac{1}{M(-s+1)}\left(\frac{1}{4}\gamma^2 - \alpha^2\right)^{\frac{1}{2}}.$ (12)

Then the solution of equation (6) has the form

$$P_0(u) = \frac{1}{u^s} e^{\frac{1}{2M(-s+1)}\gamma u^{-s+1}} e^{k(\alpha) u^{-s+1}} \tag{13}$$

Thus according to equation (2) a non-stationary distribution function due to the linearity of the Fokker-Planck equation is determined as

$$P(x,t) = \frac{e^{\frac{1}{2M(-s+1)}\gamma u^{-s+1}}}{u^s} \int_{-\infty}^{\infty} e^{k(\alpha) u^{-s+1}} e^{-\alpha^2 t}\, d\alpha \tag{14}$$

In the case of fractional time-equation of Fokker-Planck we have

$$\frac{1}{T}\frac{d^v T}{dt^v} = \frac{1}{P_0(u)}\frac{d}{du}\left[-\gamma u^s P_0(u) + M u^s \frac{d}{du}(u^s P_0(u))\right] \tag{15}$$

In this case the equation to determine the time dependence $T(t)$ is a fractional differential equation

$$\frac{d^v T(t)}{dt^v} = \omega^2 T(t) \tag{16}$$

A particular solution of this equation has the form

$$T(t) = E_v(-(\omega t)^v) \tag{17}$$

where $E_v(-(\omega t)^v)$ is the Mittag-Leffler function. In this case the solution of equation (15) has the form

$$P(x,t) = T(t)P_0(u) = E_v(-(\omega t)^v)\frac{1}{u^s} e^{\frac{1}{2M(-s+1)}\gamma u^{-s+1}} e^{k(\alpha) u^{-s+1}} \tag{18}$$

As at $v \to 1$ $\lim_{v \to 1} E_v(z) = e^z$ equation (18) coincides with solution (14).

**Conclusion**

We derive an equation of motion for the order parameter in small-world networks. The obtained equation for the order parameter is a generalization of the Landau – Ginsburg equation. With the multiplicative and additive noises taken into account the Fokker – Planck equation has been obtained. Based on the stationary solution of this equation the analysis of the induced phase transitions in such systems has been performed. The results of this paper are in good agreement with those of data processing on the dynamics of ranging processes presented in [4] Non-stationary distribution functions for scale-invariant networks with a small-world property and memory and without memory have also been defined in this paper. The obtained results can be used for the description of the dynamic processes in the scale-invariant networks with a small-world property.